\documentstyle[prb,aps,tighten,multicol,epsf]{revtex} 
 
\begin{document} 
 
\draft 
 
\title{Transport in finite incommensurate Peierls-Fr\"ohlich systems}
 
\author{Luis E. Oxman, Eduardo R. Mucciolo}

\address{Departamento de F\'{\i}sica, Pontif\'{\i}cia Universidade
Cat\'olica do Rio de Janeiro,\\ Caixa Postal 38071, 22452-970 Rio de
Janeiro, Brazil}

\author{Ilya V. Krive}

\address{B.I. Verkin Institute for Low Temperature Physics and
Engineering, Kharkov, Ukraine}

\date{May 17, 1999} 
 
\maketitle 
 

\begin{abstract}
We show that the conductance of a one-dimensional, finite
charge-density-wave (CDW) system of the incommensurate type is not
renormalized at low temperatures and depends solely on the
leads. Within our formalism, we argue that a similar behavior (perfect
conductance) should occur for a wide class of one-dimensional strongly
correlated finite systems where interactions are current
dependent. The universal conductance is related to the presence of an
(anomalous) chiral symmetry. The fundamental role played by the
finiteness of the sample and the adiabaticity of the contacts to
Fermi-liquid leads is evidenced.
\end{abstract} 

 
 

\begin{multicols}{2} 
 
\narrowtext 


\section{Introduction} 

Transport in strongly correlated, low-dimensional mesoscopic systems
has been a subject of intense investigation. The theoretical tools for
studying one-dimensional (1D) systems (exact solutions, bosonization,
renormalization group, etc.) are already quite well-known in the
literature \cite{Fradkin91} and some important concepts were developed
in the last two decades. For instance, it is understood that 1D,
interacting gapless systems at low energies fall into the Luttinger
liquid universality class.\cite{Haldane81} The thermodynamics and the
dynamical correlation functions of this liquid are also well
established. Its transport properties however are less understood. For
some time it was believed that the linear conductance of a Luttinger
liquid at zero temperature is renormalized by the interaction
strength.\cite{Apel82} Recent advances in technology allowed the
fabrication and manipulation of nanostructures, such as quantum wires,
where this theoretical prediction could be tested. The first result
found was negative,\cite{Tarucha95} namely, the measured conductance
was equal to the quantum $e^2/h$ for each propagating channel; later,
some deviation from this value was observed.\cite{Yacoby96} Soon it
was understood that, for a {\it finite} Luttinger liquid junction or
wire coupled to leads, the conductance is dominated by the
noninteracting electron gas in the leads, i.e., it should not be
renormalized by the interaction in the
wire.\cite{Maslov95,Ponomarenko95,Safi95} Why this argument fails to
explain particularly the results of Ref.~\onlinecite{Yacoby96} remains
an open issue. On the other hand, one expects that processes leading
to backscattering, such as disorder, could be
involved.\cite{Maslov95b}

In contrast to these findings, one would imagine that a junction
formed by a gapped system, like a half-filled Mott-Hubbard insulator
or a Peierls-Fr\"ohlich CDW should show an exponentially suppressed
conductance at sufficiently low temperatures. Recently, Ponomarenko
and Nagaosa \cite{Ponomarenko98a} demonstrated that a Hubbard-Mott
insulator where the order parameter (gap) is dynamical has in fact
perfect conductance, $G=2e^2/h$ (the factor of 2 stands for spin
degeneracy), i.e., transport is also controlled by the leads. In this
work we show that a similar behavior occurs for an incommensurate CDW
system adiabatically connected to Fermi-liquid leads. This result
agrees with that obtained in Ref.~\onlinecite{Rejaei96}, where
transport of charge in disordered mesoscopic CDW heterostructures was
studied within the Keldysh formalism. In contrast to the elaborate
calculation of Ref.~\onlinecite{Rejaei96}, here we present very simple
physical arguments based on the existence of a chiral symmetry for the
system as a whole (CDW plus leads) when the phase of the CDW order
parameter is dynamic. This symmetry becomes anomalous upon the
application of an external electric field. Outside the sample, the
resulting chiral current is associated to the transport of free
electrons; inside the CDW region, the current has an extra component
related to the dynamics of the lattice degrees of freedom. As long as
these degrees of freedom are not frozen or pinned by impurities, the
sample conductance will depend solely on the physics outside the CDW
region.

One is tempted to try to understand the underlying reason for such
universal behavior in a system-independent manner. We argue that
perfect conductance should be a general property of all
one-dimensional Fermi-liquid/finite-system/Fermi-liquid structures, as
long as they display a (anomalous) chiral symmetry. This occurs in
turn when the finite system possesses this symmetry and is
adiabatically connected to the Fermi-liquid reservoirs. Our work, in
some sense, makes more specific some general ideas on this subject
which have recently appeared in the literature.\cite{Alekseev98}


\section{The Peierls-Fr\"ohlich system} 

The model effective Lagrangian density that describes a 1D
incommensurate charge-density-wave system can be divided into three
terms\cite{Fukuyama85} ($\hbar = c = 1$), namely,
\begin{equation}
{\cal L}_{\mbox{\scriptsize{CDW}}} = {\cal L}_{\mbox{\scriptsize{el}}}
+ {\cal L}_{\mbox{\scriptsize{el-ph}}} + {\cal
L}_{\mbox{\scriptsize{ph}}},
\label{eq:CDWL}
\end{equation}
where
\begin{equation}
{\cal L}_{\mbox{\scriptsize{el}}}[\psi] = i\bar{\psi} \left(
\partial_\mu + ieA_\mu \right) \gamma^\mu \psi,
\end{equation}
\begin{equation}
{\cal L}_{\mbox{\scriptsize{el-ph}}}[\psi,\Delta] = \Delta \bar{\psi}
P_L \psi + \Delta^\dagger \bar{\psi} P_R \psi,
\end{equation}
and
\begin{equation}
{\cal L}_{\mbox{\scriptsize{ph}}}[\Delta] = \frac{1}{2v} \left(
\partial_t \Delta^\dagger \partial_t \Delta - v^2 \partial_x
\Delta^\dagger \partial_x \Delta \right) - \frac{\omega_{p}^2}{2v}
\Delta^\dagger \Delta.
\end{equation}
Let us specify our conventions. We follow the usual space-time
notation $x^\mu \equiv (x^0,x^1) = (t,x)$ and $\partial_\mu \equiv
\partial/\partial x^\mu$, with the metric tensor $g_{\mu\nu} =
\mbox{diag}(1,-1)$. $v$ denotes the CDW velocity in units such that
the Fermi velocity $v_F=1$. The Dirac fermionic field $\psi$ has
right- and left-moving components (the electron modes), $ \psi =
\left( \psi_L,\psi_R \right)^T$, spanning a two-dimensional vector
space where the following gamma matrices act: $\gamma_0 \equiv
\sigma_1$, $\gamma_1 \equiv -i\sigma_2$, and $\gamma_5 \equiv \gamma_0
\gamma_1 = \sigma_3$. Here, $\sigma_1,\sigma_2,\sigma_3$ are the
standard Pauli matrices. We also have $\bar{\psi} \equiv \psi^\dagger
\gamma_0$, whereas the right- and left-moving projectors are defined
as $P_{L,R} \equiv (1\pm\gamma_5)/2$. (For the sake of simplicity,
throughout this work we assume that the fermions are spinless.) The
field $\Delta$ and its adjoint $\Delta^\dagger$ represent lattice
harmonic, optical phonon modes of frequency $\omega_{p}$ and momenta
$\pm 2p_F$, where $p_F$ is the Fermi momentum. By assumption, $h/p_F$
is incommensurate with the lattice constant.

In order to implement the adiabatic transition between the
Peierls-Fr\"ohlich system (hereafter named junction) and the
asymptotic Fermi-liquid leads we replace $\Delta(x,t)$ by
$h(x)\Delta(x,t)$, where $h(x)$ is a smooth function that vanishes
outside a region $\Omega$ of length $L$; inside this region, $h(x)$
jumps to the saturation value 1. This choice allows us to describe the
complete system (junction plus leads) with a single Lagrangian
density. Problems related to wave function matching at the interfaces
are thus avoided.

When the field $\Delta$ is frozen, chiral symmetry is explicitly
broken even at the classical level and the fermions acquire a
mass. The gap in the spectrum of excitations yields an insulating
behavior and charge transport through the junction region is
exponentially suppressed. For $T\ll\Delta$, an elementary calculation
yields the electric conductance
\begin{equation}
G = \frac{4e^2}{h} \exp \left( -\frac{2\Delta} {\Delta_L} \right),
\label{eq:Gsmall}
\end{equation}
where $\Delta_L \equiv \hbar v_F/L \ll \Delta$. Notice that
exponentially small conductances are also the prediction of a
quasiclassical theory of normal-metal/CDW junctions developed in
Ref.~\onlinecite{Visscher96}. If, on the other hand, we allow for
$\Delta$ to be a dynamical field, classical global chiral symmetry is
restored. We show below that, in this situation, charge transport in
the CDW junction is strongly enhanced.


\section{Bosonization} 

Let us bosonize the fermionic fields following the standard procedure,
which includes the chiral anomaly in a consistent way. The universal
rules are\cite{Coleman75}
\begin{equation}
i \bar{\psi} \partial_\mu \gamma^\mu \psi \leftrightarrow \frac{1}{2}
\partial_\mu \phi \partial^\mu \phi,
\end{equation}
\begin{equation}
j^{\mbox{\scriptsize{V}}}_\mu \equiv \bar{\psi} \gamma^\mu \psi
\leftrightarrow \frac{\beta}{2\pi} \epsilon_{\mu\nu} \partial^\nu
\phi,
\qquad
j^{\mbox{\scriptsize{A}}}_\mu \equiv \bar{\psi} \gamma^\mu
\gamma^5\psi \leftrightarrow \frac{\beta}{2\pi} \partial^\mu \phi,
\end{equation}
\begin{equation}
\bar{\psi} P_{L} \psi \leftrightarrow \frac{C}{2} e^{i\beta\phi},
\qquad
\bar{\psi} P_{R} \psi \leftrightarrow \frac{C^\ast}{2} e^{-i\beta\phi},
\end{equation}
where $\beta = 2\sqrt{\pi}$ and $\phi$ is a massless boson field. The
constant $C$ depends on how the theory (path integral) is
renormalized. After the rules stated above, the bosonized Lagrangian
density becomes
\begin{eqnarray}
{\cal L}_{\mbox{\scriptsize{boson}}} & = & \frac{1}{2} \partial_\mu
\phi \partial^\mu \phi + \frac{h}{2} \left( C^\ast \Delta^\dagger
e^{-i\beta\phi} + C \Delta e^{i\beta\phi} \right) \nonumber \\ & & +
\frac{e\beta}{2\pi} \epsilon^{\mu\nu} \partial_\mu A_\nu \phi
\nonumber + \frac{h^2}{2v} \partial_t \Delta^\dagger \partial_t \Delta
\nonumber \\ & & - \frac{v}{2} \partial_x \left( h \Delta^\dagger
\right) \partial_x \left( h\Delta \right) - \frac{\omega_{p}^2
h^2}{2v} \Delta^\dagger \Delta.
\label{eq:bCDWL}
\end{eqnarray}
The equations of motion for $\phi$ and $\Delta$ are, respectively,
\begin{equation}
\partial_\mu \partial^\mu \phi + \frac{i\beta h}{2} \left( C^\ast
\Delta^\dagger e^{-i\beta\phi} - C \Delta e^{i\beta\phi} \right) =
-\frac{e\beta}{2\pi} E(x,t)
\label{eq:motionphi}
\end{equation}
and
\begin{equation}
\frac{1}{2v} h^2 \partial^2_t \Delta - \frac{vh}{2} \partial^2_x (h
\Delta ) + \frac{\omega_{p}^2 h^2}{2v} \Delta = - \frac{C^\ast h}{2}
e^{-i\beta\phi},
\label{eq:motionDelta}
\end{equation}
where $E(x,t) = \partial_t A_1(x,t) - \partial_x A_0(x,t)$. If we
multiply Eq.~(\ref{eq:motionDelta}) by $\Delta^\dagger$ and subtract
from the resulting expression its adjoint, we find, after using
Eq.~(\ref{eq:motionphi}), that
\begin{equation}
\partial_\mu j^\mu_{\mbox{\scriptsize{A}}} = - \frac{e\beta}{2\pi}
E(x,t),
\label{eq:chiralanomaly}
\end{equation}
where the total axial current density components are
\begin{equation}
j^{\mbox{\scriptsize{A}}}_0 = \frac{i\beta h^2}{2v} \left(
\Delta^\dagger \partial_t \Delta \right) - \left( \Delta \partial_t
\Delta^\dagger \right) + \partial_t \phi
\end{equation}
and
\begin{equation}
j^{\mbox{\scriptsize{A}}}_1 = \frac{i\beta v}{2} \left[ \left( h
\Delta^\dagger \right) \partial_x \left( h\Delta \right) - \partial_x
\left( h \Delta^\dagger \right) \left( h\Delta \right) \right] +
\partial_x \phi.
\end{equation}
Equation (\ref{eq:chiralanomaly}) is the anomalous divergence of the
chiral current associated to the classical global symmetry
\begin{equation}
\phi \rightarrow \phi + \theta, \qquad \Delta \rightarrow \Delta
e^{-i\beta\theta}.
\label{eq:symm}
\end{equation}
(In terms of the original fermionic fields, $\psi \rightarrow
e^{i\theta \gamma_5} \psi$). It is important to notice that the chiral
current depends on the particular region we are considering. Outside
the junction, where $h(x)=0$, the current is due to electrons
only. Inside the junction, both electrons and lattice degrees of
freedom contribute to the chiral current.


\section{Solution to the field equations} 

We can understand why the conductance does not depend on the
properties of the CDW junction by solving the equations of motion
(\ref{eq:motionphi}) and (\ref{eq:motionDelta}) for the fields $\phi$
and $\Delta$. This line of reasoning was introduced by Maslov and
Stone \cite{Maslov95} to arrive at a similar conclusion in the case of
a Luttinger liquid wire connected to Fermi-liquid reservoirs. We thus
assume that the electric field is zero outside the region $\Omega$,
which we take extending from $-L/2$ to $+L/2$. The electric field
$E(x,t)$ is switched on at an instant $t=t_{-}$ and eventually reaches
the stationary value $E(x)$. To facilitate the discussion, let us
assume that the electric field $E(x)$ and the profile $h(x)$ are even
functions of $x$. As a result, the field $\phi$ will have the same
property.

For late enough times, the charge current $I =
ej_1^{\scriptsize{\mbox{V}}} = -(e\beta/2\pi) \partial_t \phi$ should
be stationary (time independent) in the region around $\Omega$. The
(causal) solution of Eq.~(\ref{eq:motionphi}) compatible with these
constraints is $\phi(x,t) = -k(t \pm x)+\phi_0$, where the plus
(minus) sign corresponds to $x<-L/2$ ($x> L/2$) and
\begin{equation}
k = \frac{2\pi I}{e\beta}.
\label{eq:current}
\end{equation}
We now propose the following asymptotic ansatz to extend this solution
to the interior of $\Omega$ for late times:
\begin{equation}
\phi(x,t) = \phi_a(x,t) \equiv f(x) - kt
\label{eq:asymptoticphi}
\end{equation}
and
\begin{equation}
\Delta (x,t) = \Delta_a \equiv D(x) e^{i\beta kt},
\label{eq:asymptoticDelta}
\end{equation}
where the functions $f(x)$ and $D(x)$ satisfy the coupled differential
equations
\begin{equation}
\partial^2_x f + \frac{i\beta h}{2} \left( C^\ast D^\ast e^{-i\beta f}
+ C D e^{i\beta f} \right) = -\frac{e\beta}{2\pi} E(x)
\label{eq:staticf}
\end{equation}
and
\begin{equation}
\frac{h^2 \beta^2 k^2}{2v} D + \frac{v h}{2} \partial^2_x (h D ) - 
\frac{\omega_{p}^2 h^2}{2v} D  = \frac{C^\ast h}{2} e^{-i\beta f},
\label{eq:staticD}
\end{equation}
The correct large-distance asymptotics are implemented assuming that
$f(x) = -kx+\phi_0$ for $x < -L/2$ and $f(x) = kx+\phi_0$ for $x >
L/2$. The consistency of the ansatz will be checked in the next
section. For now, we remark that it is not necessary to know the exact
forms of $f(x)$ and $D(x)$ to calculate the conductance of the
junction. The reason is the following. Using the anomalous divergence
of Eq.~(\ref{eq:chiralanomaly}), together with
Eqs.~(\ref{eq:asymptoticphi}) and (\ref{eq:asymptoticDelta}), we
arrive at
\begin{equation}
\partial_x \left\{ \frac{i\beta vh}{2} \left[ D^\ast \partial_x (hD) -
D \partial_x (hD^\ast) \right] + \partial_x f \right\} =
\frac{e\beta}{2\pi} E(x).
\end{equation}
Integrating this expression between points $a$ ($a<-L/2$) and $b$
($b>L/2$) and since $h(a)=h(b)=0$ outside $\Omega$, we get
\begin{equation}
2k=\partial_x f(b) - \partial_x f(a) = \frac{e\beta}{2\pi} \left[ V(a)
- V(b) \right],
\label{eq:kval}
\end{equation}
where $V(x)$ is the electric potential. Recalling
Eq.~(\ref{eq:current}), we arrive at the relation
\begin{equation}
I = \frac{e^2}{2\pi} \left[ V(a) - V(b) \right],
\label{eq:percurr}
\end{equation}
which leads to the linear conductance (restoring the $\hbar$ factor),
\begin{equation}
G = \frac{e^2}{h}.
\end{equation}
The conductance is perfect, that is, it is not renormalized by the
dynamics inside the junction, only depending on the properties of the
electron gas in the leads. An interpretation of this result in terms
of the dynamics of the fields $\phi$ and $\Delta$ is in order. Notice
that in the asymptotic regime, the bosonized electron field and the
phonon phase ``rotate'' coherently, allowing for a perfect matching
between charged wave packets traveling in and out of the junction
region. The time evolution represented by
Eqs.~(\ref{eq:asymptoticphi}) and (\ref{eq:asymptoticDelta}) is very
similar to the chiral transformation of Eq.~(\ref{eq:symm}). Thus, the
action cost of this time evolution is rather small and favorable. The
lattice is left in an excited oscillating state, with a frequency
controlled by the bias and the Fermi velocity in the leads, while a
perfect electric current is established through the system.


\section{Asymptotic {\it in} and {\it out} behaviors} 

Notice that $f(x)$ is a continuous and differentiable function that
grows linearly for $|x|>L/2$ and is bounded for $|x|<L/2$. Therefore,
if $t>t_{+}$ and for large enough $t_{+}$, $\phi_a(x,t)$ will be
negative in the interval $[l(t),r(t)]$ and positive outside, where
$l(t) = -t + \phi_0/k <-L/2$ and $r(t) = t -\phi_0/k >L/2$. The points
$l(t)$ and $r(t)$ move in opposite directions, away from
$\Omega$. They propagate through the leads, with Fermi velocity, the
information about the switching on of the electric field in the region
$\Omega$ at earlier times. Thus, to be accurate, the asymptotic
solution for the field $\phi$ should be represented by
\begin{equation}
\phi_a(x,t) = [f(x) - kt] \theta \biglb(kt - f(x) \bigrb),
\label{eq:phifull}
\end{equation}
where $\theta(z)$ is the step function. It is easy to see by direct
substitution that $\phi_a$ and $\Delta_a$ satisfy the time dependent
equations (\ref{eq:motionphi}) and (\ref{eq:motionDelta}) with the
stabilized electric field profile $E(x)$; hence, at a given instant
$t>t_+$, the asymptotic electric current is given by
Eq.~(\ref{eq:percurr}) inside $[l(t),r(t)]$, and is zero outside.

In order to check the consistency of the asymptotic solution proposed
for the field $\phi$ with a periodic time dependence for $\Delta$, let
us solve Eq.~(\ref{eq:motionDelta}) for $\Delta(x,t)$ in terms of
$\phi(x,t)$. Inverting Eq.~(\ref{eq:motionDelta}), we can write that
\begin{eqnarray}
\Delta(x,t) & = & -\frac{C^\ast}{2} \int dx^\prime \int_{-\infty}^t
dt^\prime\ {\cal G}(x,t,x^\prime,t^\prime) h(x^\prime) \nonumber \\ &
& \times e^{ -i\beta \phi(x^\prime,t^\prime) },
\label{eq:inverseproblem}
\end{eqnarray}
where ${\cal G}$ is the retarded Green's function of the differential
operator
\begin{equation}
\left[ \frac{h^2}{2v} \partial^2_t - \frac{vh}{2} \left(
h^{\prime\prime} + 2h^\prime \partial_x + h \partial^2_x \right) +
\frac{h^2\omega_{p}^2}{2v} \right].
\label{eq:opera}
\end{equation}
The use of the retarded prescription is justified as follows. The
effective action obtained by path integrating over the phonon degrees
of freedom in Eq.~(\ref{eq:bCDWL}) depends explicitly on the Feynman
phonon propagator; therefore, the associated effective field equation
for the bosonized $\phi$ field is neither real nor causal.  This
happens because this equation refers to {\it in-out} vacuum
expectation values of the $\phi$ field. Alternatively, one can use a
close-time path formalism \cite{Schwinger61} to construct real and
causal effective equations for the {\it in-in} vacuum expectation
values of the Hermitian field $\phi$. It can be shown that the {\it
in-in} effective equations can be obtained by replacing, in the
corresponding {\it in-out} expressions, the Feynman propagator by the
retarded one. The resulting effective equations are equivalent to
Eqs.~(\ref{eq:motionphi}) and (\ref{eq:motionDelta}) with $\Delta$
solved by means of a retarded prescription
[c.f. Eq.~(\ref{eq:inverseproblem})].

Since the operator (\ref{eq:opera}) is time independent, it is useful
to pass to a Fourier representation, namely,
\begin{equation}
{\cal G} = {\cal G}(x,x^\prime;t-t^\prime) = \int \frac{d\omega}{2\pi}
{\cal G}_\omega(x,x^\prime) e^{-i\omega (t-t^\prime)}.
\label{eq:fourier}
\end{equation}
${\cal G}_\omega(x,x^\prime)$ will be an analytic function with poles
located in the lower half plane, but with a vanishingly small negative
imaginary part. For instance, for the limiting case $h(x)=
\theta(L/2-|x|)$, the poles are placed at $\omega_n= \pm
\sqrt{\omega_{p}^2 +v^2 \kappa_n^2}-i\epsilon$, where $\kappa_n
=(2n+1)\pi/L$, $n = 1,2,\ldots$. This pole structure implies that, for
large $t$, the main contribution to Eq.~(\ref{eq:inverseproblem})
comes from large $t^\prime$. Thus, the asymptotic behavior of
$\Delta(t)$ can be obtained by inserting in
Eq.~(\ref{eq:inverseproblem}) the asymptotic solution of
Eq.~(\ref{eq:phifull}), which in turn can be simplified to
$\phi(x^\prime,t^\prime) = f(x^\prime) -kt^\prime$, as $x^\prime$ is
limited to a finite interval. Integrating in $t^\prime$, we get the
asymptotic $\Delta$ behavior proposed in
Eq.~(\ref{eq:asymptoticDelta}).

We will now look for a time-dependent electric field $E(x,t)$ such
that $\phi=\phi_a$ for {\it all} times. For this purpose, we extend
$\phi_a$ backwards in time and try to check if the induced functions
$\Delta(x,t)$ and $E(x,t)$ represent a sensible switching on process.

For early enough times $t<t_{-}$, $f(x)-kt$ will be positive
everywhere, implying that $\phi_a(x,t)=0$. Inserting this result into
Eq.~(\ref{eq:motionphi}) yields
\begin{equation}
E(x,t) = - \frac{i\pi h}{e} \left( C^\ast \Delta^\ast - C\Delta
\right).
\label{eq:Efield}
\end{equation}
On the other hand, the retarded prescription in
Eq.~(\ref{eq:inverseproblem}) shows that at $t<t_{-}$, the integrand
should be evaluated at $t^\prime <t<t_{-}$, i.e., when $\phi = 0$:
\begin{equation}
\label{eq:solDelta}
\Delta(x,t) = -\frac{C^\ast}{2} \int dx^\prime \int_{-\infty}^0
d\tau^\prime\ {\cal G}(x,x^\prime;-\tau^\prime) h(x^\prime).
\end{equation}
We see that at early times $\Delta$ is time-independent. Since the
retarded Green's function is real, the phase of $\Delta$ is uniform
and equal to that of $C^\ast$. Inserting Eq.~(\ref{eq:solDelta}) into
Eq.~(\ref{eq:Efield}), we conclude that the induced value of the
electric field at early times ($t<t_{-}$) is zero, representing
therefore a realistic switching on process.


\section{Universality of free bosonization rules and Landauer 
conductance} 

We will now consider a generic finite system with current
interactions, adiabatically connected to Fermi-liquid leads; we will
follow the path integral approach to bosonization (see, e.g,
Ref.~\onlinecite{Roskies81}), which is based on Fujikawa's anomalous
determinant.\cite{Fujikawa79}
 
Firstly, using the method of Ref.~\onlinecite{Barci99}, we can easily
check the universality of the free bosonization rules for our
system. The fermionic partition function is
\begin{eqnarray}
Z_F[A] & = & \int {\cal D}\psi {\cal D}\bar{\psi} \exp \Bigg\{ i\int
d^2x \left[ \bar{\psi} (i\partial_\mu - eA_\mu )\gamma^\mu \psi
\right] \nonumber \\ & & + i I[j^\mu] \Bigg\}.
\label{eq:bpi}
\end{eqnarray}
The functional $I[j^\mu]$ denotes a current-dependent interaction
term, which we suppose to be localized in $\Omega$. This interaction
part can be represented in terms of a functional Fourier transform,
\begin{eqnarray}
\exp \left\{ iI[j^\mu] \right\} & = & {\cal N} \int {\cal D}a_\mu \exp
 \Bigg\{ -i\int d^2x\ h(x) a_\mu j^\mu \nonumber \\ & & + iS[a_\mu]
 \Bigg\}.
\label{eq:FTr}
\end{eqnarray}
The constant ${\cal N}$ is chosen such that $I[0]=0$. Defining $h(x)$
as in the previous sections, we see that, by construction,
$I[j^\mu]=0$ for currents localized outside $\Omega$. As before, the
role of the smooth function $h(x)$ is to implement the adiabatic
contact of the interacting finite system to the noninteracting leads.

We can insert Eq.~(\ref{eq:FTr}) into Eq.~(\ref{eq:bpi}) and bosonize
the fermions as in a free theory with an external source $eA_\mu +
h(x)a_\mu$,
\begin{eqnarray}
Z_F[A] & = & \int {\cal D}\phi {\cal D}a_\mu \exp \left\{ i \int d^2x
\left[ \frac{1}{2}\partial_\mu \phi \partial^ \mu \phi
\right. \right. \nonumber \\ & & \left. \left. -(eA_\mu + h
a_\mu)\frac{1}{\sqrt{\pi}}\epsilon^{\mu \nu} \partial_\nu \phi \right]
+ iS[a_\mu] \right\}.
\end{eqnarray}
Integrating over $a_\mu$, we obtain the bosonized action
\begin{eqnarray}
S_{\scriptsize{\mbox{boson}}} & = & \int d^2x \left(
\frac{1}{2}\partial_\mu \phi \partial^ \mu \phi - eA_\mu
\frac{1}{\sqrt{\pi}}\epsilon^{\mu \nu} \partial_\nu \phi \right)
\nonumber \\ & & + I\left[\frac{1}{\sqrt{\pi}}\epsilon^{\mu \nu}
\partial_\nu \phi \right].
\label{eq:simb}
\end{eqnarray}
This is the universality of the free bosonization rules; it implies
the universality of Landauer conductance at $T=0$ for a finite system
with current interactions localized in the junction. Indeed, the
bosonized equation of motion corresponding to Eq.~(\ref{eq:simb}) can
be written as an anomalous divergence,
\begin{equation}
\partial_\mu \left[ \partial^\mu \phi +
\frac{1}{\sqrt{\pi}}\epsilon^{\mu \nu} \frac{\delta I}{\delta
j^\nu(x)} \right] = -\frac{e\beta}{2\pi} E(x,t).
\label{eq:genan}
\end{equation}
Here, the asymptotic behavior will also have the form of
Eq.~(\ref{eq:phifull}), but with $f(x)$ replacing $\phi$ in
Eq.~(\ref{eq:genan}) for a stationary electric field $E(x)$. Notice
that the contribution to the axial current coming from the interaction
is localized. Therefore, the integration of Eq.~(\ref{eq:genan}) over
the spatial coordinate, in analogy to Eq.~(\ref{eq:kval}), will lead
to a perfect conductance again.

An analogous result is obtained if we consider a finite CDW system
with an additional (local) current interaction of the form shown in
Eq.~(\ref{eq:FTr}). We can follow the same steps that led from
Eq.~(\ref{eq:bpi}) to Eq.~(\ref{eq:simb}). The only difference is that
we must replace the massless free fermionic part in Eq.~(\ref{eq:bpi})
by the corresponding CDW Lagrangian of Eq.~(\ref{eq:CDWL}) and, of
course, include the path integration over the dynamical field
$\Delta$. The equivalence between the partition function for free
fermions, with mass parameter $\Delta (x)$, and the corresponding
bosonized version\cite{Naon85} can then be used to complete the steps
showing the universality of the free bosonization rules. When the
bosonized interaction term $I[1/\sqrt{\pi}\, \epsilon^{\mu \nu}
\partial_\nu \phi]$ is added to the bosonized CDW action
[c.f. Eq.~(\ref{eq:bCDWL})], the anomalous chiral current will be
modified by a term localized in the junction region. This kind of
term, as we have seen, does not change the transport properties of the
system. For instance, these conclusions hold when forward-scattering
impurities are present in the CDW junction; if impurities are also
present in the Fermi-liquid leads, however, some renormalization of
the conductance is expected, in agreement with the results of
Ref.~\onlinecite{Rejaei96}.


\section{Summary}

We have shown that one-dimensional structures of the
Fermi-liquid/finite-system/Fermi-liquid type display perfect Landauer
conductance quantization at low temperatures, provided the finite
system part presents an anomalous chiral symmetry. An important
ingredient behind this behavior is the adiabaticity of the contacts
between the finite system and the Fermi-liquid leads or reservoirs,
which permits the extension of the chiral symmetry to the system as a
whole. In this case, an anomalous chiral current is always present
when a bias voltage is applied; outside the sample, this current is
associated to the transport of free fermions. These general properties
cause the charge transport through the system to be dominated by the
reservoirs.

The natural language used to study this problems is bosonization. In
$1+1$ dimensions, it offers a simple way to take into account the
chiral anomaly. In this framework, we have shown that the universality
of Landauer conductance is intimately connected to the universality of
the free bosonization rules.

The general relation between universality in transport properties of
gapless systems and the existence of conserved chiral charges was
first proposed in Ref.~\onlinecite{Alekseev98}. In that work, however,
the role of finite sample size and the adiabaticity of the contacts
was not apparent. We hope to have contributed to clarify this point
with our discussion. In particular, we have studied a finite CDW
system of the incommensurate type. The dynamical character of the
phonon field is responsible for the restoration of the anomalous
chiral symmetry, which is absent in the case of static phonon
fields. This system, when adiabatically connected to reservoirs,
presents a nonrenormalized, Fermi-liquid like conductance at low
temperatures.

We estimate that this conductance quantization should be
experimentally observable up to temperatures $k_BT$ of order
$\hbar\omega_{p}$ (the optical phonon frequency). For higher
temperatures, fluctuations could destroy the CDW phase coherence. In
that case, the conductance would involve a thermal activation
mechanism, as in a band gap material. We have seen that the addition
of forward scattering impurities to the CDW junction will not modify
its perfect conductance. On the other hand, it is important to remark
that, in practice, any strong inhomogeneities, leading to
backscattering, will break global chiral symmetry. In a similar way,
the pinning of the phonon field phase by impurities will cause the
conductance to be suppressed. The effect of commensurability of the
CDW wavelength with the underlying lattice constant will be discussed
elsewhere.\cite{commensurate}

\acknowledgments

This work was partially supported by the Brazilian agencies CAPES and
CNPq and the NRC's Twinning Program for Ukraine
(1999-2000). I.K. thanks the Physics Department at PUC-Rio (Rio de
Janeiro) and the Department of Applied Physics at Chalmers University
of Technology (G\"oteborg) for the hospitality, as well as
A. S. Rozhavsky and R. I. Shekhter for fruitful discussions.



\end{multicols} 

\end{document}